\documentstyle[12pt]{article}
\def\deg{\ifmmode{^{\circ}}\else ${^{\circ}}$\fi}

\def\ni#1{\noindent$(#1)\quad$}

\newcommand{\bi}{\begin{itemize}}
\newcommand{\ei}{\end{itemize}}
\def\ed{\end{document}}
\begin{document}
\newcommand{\lqcd  }{ \mbox{$\Lambda_{QCD} $} }
\newcommand{\lqcdsq  }{ \mbox{$\Lambda^2_{QCD} $} }
\begin{titlepage}
\begin{flushright}
{\sl NUB-3061/93-Th}\\
hep-ph/9304238\\
{\sl April 1992}
\end{flushright}
\vskip 0.5in
\begin{center}
{\Large\bf Large Multiplicity Gluon Production in QCD }\\[.5in]
{\sc Haim Goldberg}\footnote{email: goldberg@neuhep.hex.northeastern.edu}
{\sc and Rogerio Rosenfeld}\footnote{email:
rosenfeld@neuhep.hex.northeastern.edu}
\\[.5in]
{\sl Department of Physics}\\
{\sl Northeastern University}\\
{\sl Boston, MA 02115}
\end{center}
\vskip 1in
\begin{abstract}
We compute the approximate cross section $\sigma_n^{(0)}$ for
producing $n-2$ {\it resolved} gluons in a
gluon-gluon collision, using
the Parke-Taylor formula regularized in a Lorentz invariant manner.
We find, in double leading logarithm approximation, that
\smallskip

\[
\sigma_{n}^{(0)} \approx \frac{1}{s}  \left( \frac{N_c\alpha_s}{2 \pi
\sqrt{12}}
                    \ \ln^{2} (s/s_{cut})\right)^{n-2}\ \ ,
\]\smallskip

\noindent
where $\sqrt{s_{cut}}$ is the minimum invariant mass for a resolved gluon pair.
There is no factor of $1/(n-2)!$ multiplying the expression. We present
additional numerical results, and comment on their
implications for perturbative calculations of $n$-jet cross
sections at colliders.
\end{abstract}
\end{titlepage}
\clearpage
\setcounter{page}{2}
\section*{Introduction}
There is little doubt that QCD, the $SU(3)$ gauge theory describing
the interactions of quarks and gluons, is the underlying theory of strong
interactions.
Various experimental results from both electron-positron
machines \cite{e+e-} and
hadron colliders \cite{hadron} are in good agreement with QCD predictions.
Use of sophisticated Monte Carlo generators based on coherent
branching processes which can take into account both initial and final state
radiation \cite{webber} are able to reproduce the CDF data \cite{CDF}
up to six jets.

Multiple QCD jets will, of course, constitute important background for new
physics discovery at the future SSC/LHC colliders. Thus, there is practical
importance in being able to calculate differential cross sections for $n$-jet
processes in QCD at collider energies, with cuts appropriate to experiments
and detectors of interest. At a more theoretical level, recent work on the
behavior of multiparticle amplitudes in  field theories of massive bosons
\cite{cornwall,goldberg,voloshin,argyres,brown} has revealed non-perturbative
behavior for tree level cross sections near threshold. It is an interesting
question whether the most established bosonic field theory (QCD) maintains a
perturbative behavior at tree level for multiple jet production.

In this Letter, we will focus on the behavior of the {\it exclusive} cross
section for the process $gg\rightarrow (n-2)\ g\ $\cite{realistic}.
All final state partons will be separated in a Lorentz-invariant way, by a
given fixed minimum amount $(p_i+p_j)^2\geq s_{cut}.$
We will dispense with regions of phase space where
our cuts are not obeyed, so that we do not merge unresolved gluons into jets;
it is in this sense that our result will be an exclusive cross section. We will
concentrate on the limiting case of the production of a  large number of final
state gluons. Our results will depend on the ratio  $\Delta = s_{cut}/s$,
where $s$ is the parton-parton center-of-mass (c.m.)  energy squared (usually
denoted by $\hat s$).  Different values of $\Delta$ discriminate whether or not
the produced jets constitute important backgrounds for new physics; it is also
possible that multi-minijet jet cross sections (small $\Delta$)
may be the most important QCD contribution to
the  total hadronic cross section. A typical value of $\Delta$ for interesting
jets at the SSC ($p_{Ti} > 50\   \mbox{GeV},\  \theta_{ij} > 30^{\circ},\
\sqrt{s} \approx 4 \ \mbox{TeV} )$  is $\Delta \approx 10^{-5}$.

We will also be concerned with the validity of perturbation theory
in the approximate tree-level calculation we will perform.
Naively, each time there is an extra parton in the final state one should
pay a price of a coupling constant.
However, it is possible that in calculating certain quantities the
expansion parameter becomes effectively the coupling constant multiplied by
a number that may be large. In this case, care must be taken either by
summing up these large contributions to all orders or, more modestly,
performing perturbative calculations only when the effective expansion
parameter
is small.

Exact multi-parton QCD amplitudes may be generated through use of the
Berends-Giele recursion relations \cite{BG}; amplitudes for
$n=8$ (six final state gluons) have been given explicitly \cite{BGK}.
However, the complexity of these amplitudes  makes
their usefulness rather limited by the computer time required for their
evaluation.
Since it is the aim of this work to examine cross sections for large
numbers of final state gluons, we will search for
a reliable approximation for these exact multiparticle cross
sections.
\section*{The Parke-Taylor Formula}
As a first step to such an approximation, we will use as the  squared
amplitude for the process $gg\rightarrow (n-2)\ g$  the so-called Parke-Taylor
(PT) formula \cite{PT}, which was conjectured in
Ref.~\cite{PT},
and
later proved in
Ref.~\cite{BG}.
Quite surprisingly,
the PT result can also be derived from
soft-gluon factorization techniques \cite{Fiorani}.
This formula is the result in leading order in the
number of colors $(N_{c} = 3)\ $ for the square of the amplitude describing
the process $gg \rightarrow (n-2) g$, summed over colors and over a
particular set of gluon helicity configurations :
\begin{equation}
|{\cal A}_{n}^{PT}|^{2} = 2 g_{s}^{2n-4} N_{c}^{n-2} (N_{c}^{2} - 1) \;
\sum_{i > j} s_{ij}^{4}  \sum_{ P'(1,2,...,n)} \frac{1}{
s_{12} s_{23} s_{34} ... s_{n1}  }
\end{equation}
where $g_{s}$ is the QCD coupling constant, $s_{ij} = (p_{i} + p_{j})^{2} =
2 p_{i} \cdot p_{j}$ and the primed sum is over the $(n-1) !/2$
non-cyclic permutations of $(1,2,...,n)$ .
The relevant set of configurations is the one where all gluons but two have
the same helicities, the so-called maximally helicity-violating (MHV)
amplitudes. There are $n (n-1)$ MHV amplitudes for $n > 4$.

This leading $N_{c}$ result contains all the leading collinear and soft
divergencies in it and furthermore it preserves all the coherence effects
and includes initial and final state radiation interference.

In comparing the PT formula with exact results one should
devise methods to take into account the contributions from amplitudes
other than MHV ones.
Here we'll adopt the simplest approach, due to Kunszt and Stirling \cite{KS}
which is to assume that all the amplitudes are of comparable size. We'll call
this procedure the KS approximation.
Hence, since there are $2^{n} - 2 (n+1)$
non-zero helicity amplitudes in total we
simply multiply the PT formula by a combinatorial factor which
counts the number of non-vanishing amplitudes:
\begin{equation}
|{\cal A}_{n}^{KS}|^{2} = \frac{2^{n} - 2 (n+1)}{n (n-1)}
|{\cal A}_{n}^{PT}|^{2} .
\label{eq:ks}
\end{equation}

Several groups have performed comparisons between the PT formula
and the results from exact calculations. In order to find finite cross
sections, a certain set of cuts must be imposed on the final state parton
momenta.
Usually one is interested in large transverse momentum jets which may turn
out to be backgrounds for interesting new physics. Typical cuts are :
$ p_{Ti} \geq 50\ \mbox{GeV} \;\; ; \;\; |\eta_{i}| \leq 3  \;\; ; \;\;
\theta_{ij} \geq 30^{\circ} $ for the LHC.
Kunszt and Stirling \cite{KS} have found good agreement between the exact
$gg \rightarrow gggg$ cross section and the PT formula in the KS
approximation and they have also used the PT result to compute up to
$gg \rightarrow gggggg$ and found a broad agreement with the multiplicity of
large $p_{T}$ jets up to six jets seen at the UA1 detector.
The exact $gg \rightarrow ggggg$ result has also been tested against
various approximations schemes \cite{BGK,KK,Maxwell} for different sets of
cuts.
The overall conclusion is that the PT formula in the KS approximation
overestimates the exact results by as much as $50 \%$ for tight cuts but
becomes a better approximation for looser cuts. This can be understood since
for loose cuts the event rate is dominated by the infrared and collinear
singularities which are taken into account by the PT formula.
Since all the gluons produced are supposed to be well separated in these
calculations, the merging of two or more gluons into jets was not
taken into account.

\section*{Phase Space Parametrization}

The form in which the PT formula is written suggests a parametrization of
the $(n-2)$-body massless phase space in terms of the variables
$s_{i,\ i+1} = (p_{i} + p_{i+1})^{2}  $\cite{BK} :\\
\begin{eqnarray}
\int d \; (PS) &=& \frac{(4 \pi)^{-2 n + 7}  }{2} \frac{1}{s} \;
\int_{0}^{s} \frac{d M_{n-3}^{2}}{s - M_{n-3}^{2}}
\int_{0}^{s^{+}_{n-2,\ n-1}} d s_{n-2,\ n-1}  \; \times \; \nonumber \\[.25in]
& &
\int_{0}^{M_{n-3}^{2}} \frac{d M_{n-4}^{2}}{ M_{n-3}^{2} - M_{n-4}^{2}}
\int_{0}^{s^{+}_{n-3,\ n-2}} d s_{n-3,\ n-2}  \; \times \;
\cdots \; \times    \nonumber \\[.25in]
& &
\int_{0}^{M_{3}^{2}} \frac{d M_{2}^{2}}{M_{3}^{2} - M_{2}^{2} }
\int_{0}^{s^{+}_{3 \: 4}} d s_{3 \: 4}  \; \times \;
\int_{0}^{s^{+}_{2 \: 3}} d s_{2 \: 3}
\end{eqnarray}\\
where $M_{i}^{2} = (p_{1} + p_{2} + \cdots + p_{i} )^2$,
$M_{n-2}^{2} = s$ , $M_{n-1}^{2} = M_{1}^{2} = 0$
 and the limits of integration on
the variables $s_{i,\ i+1}$ are given by :
\begin{equation}
s^{+}_{i,\ i+1} = \frac{1}{M_{i}^{2}} |M_{i+1}^{2} -M_{i}^{2} |
\: (M_{i}^{2} - M_{i-1}^{2}) .
\end{equation}
There are $(n-3)$ independent $s_{i,\ i+1}$ variables, $(n-4)$ independent
$M_{i}^{2}$ variables and the integration over the remaining $(n-3)$
azimuthal angles was already performed.
\clearpage

We can write
down schematically the cross section for the process $gg \rightarrow (n-2) g $
:
\begin{equation}
\sigma_{n} = \frac{1}{2 s} \int d (PS)_{n-2}  \; | {\cal A}_{n}^{KS} |^{2}
\end{equation}
In order to avoid collinear and infrared divergencies we introduce
a Lorentz invariant cut-off by requiring $s_{i \:j} = ( p_{i} + p_{j} )^{2}
\geq s_{cut} $ for all particles involved. In particular, we will not
consider merging processes, i.e., contributions to the $n$-gluon cross
section from processes with
a larger number of gluons where one or more gluons fail to pass our cut.

We are interested in the large $n$ behavior of this cross section.
Due to the factorially growing number of permutations entering in
the calculation of the PT formula and the large dimensionality of phase
space, we make two approximations that we feel are plausible
in the large-$n$ limit :
\bi
\item \underline{Ordering Independence}: For large $n,$ the symmetry of the PT
formula and of our cuts
suggests that each term makes a similar contribution to the cross
section after integrating over phase space \cite{nfac}.
It is the presence of the two fixed
initial momenta which violates this hypothesis, and it is plausible that the
effect of these momenta is less important at large $n.$
In fact, we have verified this approximation
numerically for several arbitrary permutations of the momenta, in the cases
$n=10,\ n=12$,
using the phase space generators GENBOD and RAMBO.
This simplifies our calculation
tremendously since we now have to compute the contribution from the
basic string only.
\item \underline{Neglect of end point effects}: there are $(n-3)$ independent
variables $s_{i,\ i+1}$ parametrizing the $(n-2)$-body phase space
whereas there are $n$ such variables appearing in a given string in the PT
formula. It is very difficult to write out the $3$ dependent quantities in
term of the independent variables. Since the number of dependent variables
is $n$ independent, we expect that in the large-$n$ limit their contribution
is not relevant. Therefore, here we assume that we can conservatively
substitute
$ s^{3}$ for the dependent variables.
\end{itemize}

\noindent In addition to these, we also approximate the numerator of the PT
formula by $(\sum_{i > j} s_{ij}^{4} ) \approx s^{4}.$ This we call
\underline{$s^4$ dominance}.

In the context of these approximations and with the above cuts we find :
\begin{eqnarray}
\sigma_{gg \rightarrow (n-2) g} &=& \frac{1}{2 s} \cdot
\frac{1}{4 (N_{c}^{2}-1)^{2}} \cdot \frac{1}{(n-2) !}  \cdot
\frac{2^{n} - 2 (n+1)}{n (n-1)}  \cdot \frac{ (4 \pi)^{-2 n +7}}{2 s} \cdot
  \nonumber  \\[.25in]
& &
(2 g_{s}^{2n-4} N_{c}^{n-2} (N_{c}^{2}-1) ) \cdot  s^4 \cdot (n-1)! \cdot
\frac{1}{s^{3} } \cdot  I_{n-2}(s_{cut}/s) \label{eq:cross}
\end{eqnarray}\\
where\\
\begin{eqnarray}
I_{n-2}(s_{cut}/s) &= &
\int_{ (M_{n-3}^{2})_{min}}^{(M_{n-3}^{2})_{max}}
\frac{d M_{n-3}^{2}}{s - M_{n-3}^{2}}
\int_{(M_{n-4}^{2})_{min}}^{(M_{n-4}^{2})_{max}}
\frac{d M_{n-4}^{2}}{ M_{n-3}^{2} - M_{n-4}^{2}} \cdots
\int_{(M_{2}^{2})_{min}}^{(M_{2}^{2})_{max}}
\frac{d M_{2}^{2}}{M_{3}^{2} - M_{2}^{2} } \; \times   \nonumber   \\[.25in]
& &
\int_{s_{cut}}^{s^{+}_{n-2,\ n-1}} \frac{d s_{n-2,\  n-1}}{s_{n-2,\  n-1}}
\int_{s_{cut}}^{s^{+}_{n-3,\ n-2}} \frac{d s_{n-3,\ n-2}}{s_{n-3,\  n-2}}
\cdots
\int_{s_{cut}}^{s^{+}_{3 \: 4}} \frac{d s_{3 \: 4}}{ s_{3 \: 4}}
\int_{s_{cut}}^{s^{+}_{2 \: 3}} \frac{d s_{2 \: 3}}{ s_{2 \: 3}}
\label{eq:integral1}
\end{eqnarray}\\
with
$(M_{i}^{2})_{min} = \frac{i \: !}{2 (i-2)\: ! } \: s_{cut}$ and
$(M_{i}^{2})_{max} = M_{i+1}^{2} - i \: s_{cut} $ .
We wrote out explicitely the different factors entering in
Eq.~(\ref{eq:cross}) in order to point out their origin.
Respectively these factors account for : the usual flux factor,
the average over helicities and color of the initial gluons, the
Bose symmetry factor for the $(n-2)$ identical
particles in the final state, the KS approximation,
the phase-space overall factor, the overall factor in the PT formula,
 the $s^{4}$ dominance, the ordering independence approximation and
the neglecting of end point effects.

Let us now focus on the estimation of the multiple integral $I_{n-2}( \Delta =
s_{cut}/s )$ given by Eq.~(\ref{eq:integral1}), which after a trivial
integration over the $ s_{i,\ i+1} $ variables can be written as~:
\begin{eqnarray}
I_{n-2}(\Delta) &=& \int_{(x_{1})_{min}}^{(x_{1})_{max}} \frac{d x_{1}}{1-
x_{1}}
 \int_{(x_{2})_{min}}^{(x_{2})_{max}} \frac{d x_{2}}{1- x_{2}} \cdots
 \int_{(x_{n-4})_{min}}^{(x_{n-4})_{max}} \frac{d x_{n-4}}{1- x_{n-4}} \times
\ln ( \frac{1-x_{1}}{\Delta} )
\nonumber  \\[.25in]
& & \cdot
 \ln ( \frac{(1-x_{1})(1-x_{2})}{\Delta} )
\cdot \ln ( \frac{x_{1}}{\Delta} (1-x_{2})(1-x_{3}) ) \cdot
\ln ( \frac{x_{1}}{\Delta} x_{2}(1-x_{3})(1-x_{4}) ) \cdots  \nonumber
\\[.25in]
& &
\ln ( \frac{x_{1}}{\Delta} x_{2} x_{3} \cdots x_{n-6} (1-x_{n-5}) (1-x_{n-4}) )
\cdot
\ln ( \frac{x_{1}}{\Delta} x_{2} x_{3} \cdots x_{n-5} (1-x_{n-4}) )
\label{eq:integral2}
\end{eqnarray}
where the new variables $x_{i}$ are defined by :
\begin{equation}
x_{1} = \frac{M_{n-3}^{2}}{s} \; , \; x_{2} = \frac{M_{n-4}^{2}}{s x_{1}}
\; , \; x_{3} = \frac{M_{n-5}^{2}}{s x_{1} x_{2}} \; , \ldots \; ,
 x_{n-4} = \frac{M_{2}^{2}}{s x_{1} x_{2} \cdots x_{n-6} x_{n-5}}
\end{equation}
and the limits of integration are :
\begin{equation}
(x_{i})_{max/min} = \frac{ (M_{i}^{2})_{max/min} }{s x_{1} x_{2}
\cdots x_{i-1} }
\end{equation}

At this point we could proceed by simply performing a numerical integration
of Eq.~(\ref{eq:integral2}) for different values of $\Delta = s_{cut}/s $;
this we did for a number of gluons up to
$n = 14$, and the results will be discussed shortly. However, in this section
we
would like to gain some analytic understanding of the cross section for small
enough $\Delta$ and large $n$. For these reasons, we explore ways
of finding the leading terms of $I_{n-2}(\Delta)$ in this limit.

\section*{Double Leading Log Behavior}

In order to have an approximate analytical form for $I_{n-2}(\Delta),$
we find that the
following simplifications in the integration limits are useful :
\begin{equation}
 (x_{i})_{min} = 0 \;\; , \;\;
(x_{i})_{max} = 1 - \frac{\Delta}{x_{1} x_{2} \cdots x_{i-1}}
\end{equation}
\smallskip

\begin{equation}
\int_{0}^{1 - \frac{\Delta}{x_{1} x_{2} \cdots x_{i-1}}}
\frac{d x}{1-x} \ln^{m}(1-x) \ln^{n}(x)  \approx
\int_{0}^{1}
\frac{d x}{1-x} \ln^{m}(1-x) \ln^{n}(x)  \equiv c(m,n)
\end{equation}
\smallskip

\begin{equation}
\int_{0}^{1 - \frac{\Delta}{x_{1} x_{2} \cdots x_{i-1}}}
\frac{d x}{1-x} \ln^{m}(1-x)  \approx
\int_{0}^{1 - \Delta}
\frac{d x}{1-x} \ln^{m}(1-x)   = \frac{1}{m+1} \ln^{m+1} (1/\Delta)
\end{equation}
where the numbers $c(m,n)$ can be easily computed numerically.
Within these approximations and using MATHEMATICA we are able to find
an analytical form
for $I_{n-2}(\Delta)$ for $n$ up to 11. For $n > 11$ the computer time
required becomes too large.
Nevertheless, the analytical approximation provides us with the
double leading-logarithm term (DLL) in $1/\Delta$ , $I_{n-2}^{DLL}(\Delta)$
for {\it arbitrary} $n$ :
\begin{equation}
I_{n-2}^{DLL} (\Delta) = \left\{ \begin{array}{ll}
	(\frac{1}{12})^{\frac{n-4}{2}} \ln^{2n-7} (1/\Delta)  & \mbox{ $n$ even}
\\[.25in]
\frac{1}{3}(\frac{1}{12})^{\frac{n-5}{2}} \ln^{2n-7}
(1/\Delta)  & \mbox{ $n$ odd}
			     \end{array}
			\right.    \label{eq:limit}
\end{equation}
which becomes dominant at small $\Delta$. The power of $\ln^2$ in
(\ref{eq:limit}) reflects the kinematic
constraint that not every final state gluon
momentum can
be within $s_{cut}$ of another one. This consideration becomes
unimportant at large $n.$

 We have compared the results from the numerical integration with
the approximate
analytical  and DLL results for different
values of $\Delta$ and $n$ .
We find that the analytical and numerical integration results agree over a wide
range of $n$ and $\Delta.$ For fixed $n$, our DLL approximation is valid only
at
small enough $\Delta,$ the range of validity being related to the
number of partons $n$. For our results, a rough measure of the range of
validity of the DLL is
given by :
\begin{equation}
\Delta \ll \exp [-n] \ \ .
\end{equation}

%
%which is satisfied by our results for limited $n$.

Therefore, from the above considerations we finally arrive at the asymptotic
behavior of the small $\Delta$, large
$n$-gluon exclusive cross section
by combining Eqs.~(\ref{eq:cross}) and (\ref{eq:limit}) in the large $n$
limit :\\
\begin{equation}
\sigma_{n}^{(0)} \approx \frac{1}{s}
\left( \frac{N_{c}\alpha_{s}}{2 \pi \sqrt{12}}
                                       \ln^{2} (1/\Delta) \right)^{n-2}
\label{eq:result}
\end{equation}\\
In Eq.~(\ref{eq:result}) the superscript refers to the number of
unresolved gluons. We believe this result is valid in the range
$n_0 < n \stackrel{\textstyle <}{\sim}n >  \ln (1/\Delta),$
where $n_0$ is the minimum number of gluons such that our approximations
become reliable. We don't have a definite estimate for $n_0$ but it is
not unreasonable to assume $n_0 \approx 10 $.

\section*{Results for Hard Jets}

Even at moderate values of $n$ and $\Delta$, where the DLL approximation is not
valid, our numerical results may be of relevance to SSC physics. This can
be seen by writing Eq.~(\ref{eq:cross}) as :\\
\begin{equation}
\sigma_n = \frac{1}{s} \, \left( z(n,\Delta) \right)^{n-2} \; ,
\end{equation}\\
where\\
\begin{equation}
 z(n,\Delta) = \frac{N_c\alpha_s}{2 \pi} \left( I_{n-2} (\Delta) \right)^
{\frac{1}{n-2}}  \; .
\end{equation}\\
When $ z(n,\Delta) \approx 1$ one would be suspicious of the use of
perturbation theory. In Table~1 we show $ z(n,\Delta)$ for
some interesting values of $n$ and $\Delta$, where we conservatively used
$\alpha_s(Q^2 = s_{cut}) = 0.12$.
Our results suggest that at the SSC, {\em
even reasonably ``hard'' jet cuts place
$\Delta$ at a value $(\approx 10^{-5})$
which may be approaching a threshold of uncertainty
for perturbation theory, in the sense that $\sigma^{(0)}_{n+1} >
\sigma^{(0)}_{n}.$}
In the DLL approximation,  $ z(n,\Delta)$ becomes the effective $n$-independent
expansion parameter of Eq.~\ref{eq:result}:\\
\begin{equation}
 z(n,\Delta) \stackrel{{\mbox{\small DLL}}}{\longrightarrow}
 z(\Delta) = \frac{N_c\alpha_s}{2 \pi \sqrt{12}}  \ln^2 (1/\Delta) \ \ .
\end{equation}\\

\section*{Discussion and Concluding Remarks}

\ni{1} It is not difficult to understand
how our basic result Eq.~(\ref{eq:result})
comes about: The factor $\ln^{2n} (1/\Delta)$ can be accounted for by simply
counting the integrals in Eq.~(\ref{eq:integral1}); it is a consequence of
the infrared and collinear divergences contained in the PT formula that
are regulated by $s_{cut}$. However, the  crucial feature of
Eq.~(\ref{eq:result}) is the absence of a $1/n!$ factor. This  can be
traced to the approximate ordering independence of the phase space integrated
PT formula; the different $n!$ permutations compensate for the Bose
symmetry factor.
It should be noted that this cancellation of the $n!$ does {\it not}
occur for QED: in that case, the $n!$ terms in the amplitude add to a
{\it single} term which incorporates the permutation symmetry\cite{mangano}.
\smallskip

\ni{2}The Parke-Taylor formula drops terms in the cross section which are
non-leading in $1/N_c.$ Non-leading terms in $1/N_c$ do not contribute in DLL,
\cite{MangPark}, so that we do not consider them further.
\smallskip

\ni{3} As a result of the absence of the $1/n!$ or other similar factor, the
cross sections $\sigma_n^{(0)}$ for the
production of $n-2$ resolved gluons in the DLL form a {\it geometric} series in
$z,$ and the {\it exclusive} total cross section
\[
\sigma^{(0)}_{tot} = \sum_{n} \sigma_{n}^{(0)}
\]
is not summable (except for  a phase space cutoff) for $z(\Delta)\geq 1.$
In particular, there is no sign of the exponential summation $\sigma\sim
\exp(-\frac{\alpha_s}{\pi}\ln s)$ proposed by Lipatov \cite{lipatov}.
We will comment shortly on possible
implications for {\it inclusive} $n$-jet cross sections.
\smallskip

\ni{4}The factor $\sqrt{12}$ in our small-$\Delta$ result (Eq.~\ref{eq:result})
contains the the KS approximation (Eq.~\ref{eq:ks}). If this approximation
overestimates the cross section by a factor $(1+\delta)^n,$ then the
substitution $\sqrt{12}\rightarrow \sqrt{12(1+\delta)}$ should be made in
Eq.~\ref{eq:result}.
\smallskip

\ni{5} Our results suggest that  caution should be exercised
 before using tree-level exact multi-gluon amplitudes to
compute mini-jet cross sections. When convoluting the hard cross section
with the gluon distribution functions, it is possible that (for $n$
not too large) the important contributions
to the cross section come from low values of the initial gluons'
invariant
mass $\sqrt{s}$ ({\it i.e.,} larger values of $\Delta),$
in which case the use of
perturbation theory will be valid.
However, there will always be potentially large contributions
from non-perturbative
regions of $s/s_{cut},$ and one must be certain that they are under control.
\smallskip

\ni{6}
Can higher order corrections restore the validity of perturbation theory?
Schematically, we would have to compute contributions to the {\it inclusive}
cross section $\sigma_{n}$ from processes with a
larger number of gluons where one or
more gluons would fail to pass our cuts and would be merged to
another gluon:
\begin{equation}
\sigma_{n} = \sigma^{(0)}_{n} + \sigma^{(1)}_{n+1} + \sigma^{(2)}_{n+2} +
\ldots
\end{equation}
where the superscripts indicate the number of ``unresolved'' merged gluons and
\clearpage
\begin{eqnarray}
 \sigma^{(0)}_{n} &=& A_{0} \alpha_{s}^{n} \ln^{2n} (1/\Delta)  \nonumber \\
 \sigma^{(1)}_{n+1} &=& A_{1} \alpha_{s}^{n} \ln^{2n} (1/\Delta)
\alpha_{s} (\ln^{2}(1/\epsilon) - \ln^{2}(1/\Delta) )  \nonumber  \\
\sigma^{(2)}_{n+2} &=& A_{2} \alpha_{s}^{n} \ln^{2n} (1/\Delta)
\alpha^{2}_{s} (a \ln^{4}(1/\epsilon) + b \ln^{2}(1/\epsilon) \ln^{2}(1/\Delta)
+c \ln^{4}(1/\Delta) )  \nonumber  \\
\vdots & &  \vdots
\end{eqnarray}
$\epsilon$ is an infrared cut-off whose contributions will be canceled
by loop corrections, and the various constants depend on the details of the
merging procedure. An explicit discussion in the case of $e^+e^-\rightarrow
q\bar q + \mbox{gluons}$ has been given by Brown and Stirling \cite{BS1},
and will be commented on below.  In our case,

\begin{equation}
\sigma_{n} = A_{0}\ \alpha_{s}^{n}\ \ln^{2n} (1/\Delta) \left(1 -
\frac{A_{1}}{A_{0}} \alpha_{s} \ln^{2} (1/\Delta) + \frac{A_{2}c}{A_{0}}
 \alpha_{s}^{2} \ln^{4} (1/\Delta) + \ldots\right) =
f_n(z) \sigma_{n}^{(0)}\ \ .
\end{equation}
In order that perturbation theory be restored with respect to $\sigma_n,$ it
would be necessary that $f_nz^n$ be bounded for large enough $x$ and $n.$
As an example, the usual  $n$-independent Sudakov factor,
$f_n(z) = \exp [- K z] $ ,
is {\it not} enough to solve our problem. In their study of $e^+e^-\rightarrow
\mbox{jets},$ Brown and Stirling \cite{BS1} have examined inclusive jet
production using a jet merging alogorithm similar to ours (the JADE algorithm),
and have not found evidence for the presence of Sudakov summation.
At this time, we have nothing to say about the function $f_n(z),$ but its
evaluation will be important in proceeding beyond the results of this work.

\section*{Acknowledgement}

One of us (H.G.) would like to thank Chris Maxwell for
stimulating discussions in which the idea for this project was formed, and for
equally helpful remarks on its conclusion. This research was supported in part
by the National Science Foundation under Grant No. PHY-9001439, and by the
Texas National Research Laboratory Commission under Award No. RGFY9214.

\clearpage
\begin{table}
\begin{center}
{ \renewcommand{\arraystretch}{1.1}

\begin{tabular}{|c|c|c|c|c|}   \hline
\multicolumn{1}{|c|}{$\Delta$}       &\multicolumn{1}{c|}{$n=10$}
 &\multicolumn{1}{c|}{$n=11$}   &\multicolumn{1}{c|}{$n=12$}
 &\multicolumn{1}{c|}{$n=13$}  \\  \hline
$10^{-4}$      &$0.69$  &$0.70$  &$0.70$  &$0.70$     \\  \hline
$10^{-5}$      &$1.1$   &$1.1$   &$1.1$   &$1.1$      \\ \hline
$10^{-6}$      &$1.5$   &$1.5$   &$1.6$   &$1.7$      \\ \hline
\end{tabular}   }
\end{center}
\vskip 0.5in

\caption{ Numerical values of the effective expansion parameter
$z(n,\Delta)$ for different values of
$\Delta$ and $n$ .}

\end{table}


\begin{thebibliography}{99}
\bibitem{realistic}
A realistic cross section could then be obtained by convoluting the
parton level result with the gluon distribution function inside the initial
hadrons, and with the final state  hadronization functions. This will not be
done in the present work.
\bibitem{nfac}There are in fact ${\cal O}(n!)$ terms in which the permutations
are strictly of the final state gluons. These do give identical contributions
on integration over phase space. (We would like to thank C. Maxwell for
comments
on this point.)
\end{thebibliography}
\end{document}